\documentclass[twocolumn]{aastex63}
\usepackage{amsmath}
\usepackage{textcomp}
\usepackage{multirow}
\usepackage{graphicx}
\usepackage{booktabs}
\usepackage{tabularx}
\usepackage{rotating}
\usepackage{makecell}
\usepackage[T1]{fontenc}

\newcommand{\mc}[1]{\multicolumn{1}{c}{#1}}
\newcommand{\msun}{M_\odot}
\newcommand{\mearth}{M_\oplus}

\newcolumntype{Y}{>{\centering\arraybackslash}X}

\begin{document}

\title{GW Ori: Interactions Between a Triple-star System and its Circumtriple Disk in Action}

\author{Jiaqing Bi}
\affil{Department of Physics \& Astronomy, University of Victoria, Victoria, BC V8P 5C2, Canada}
\email{jiaqing.bi@gmail.com}

\author{Nienke van der Marel}
\affil{Department of Physics \& Astronomy, University of Victoria, Victoria, BC V8P 5C2, Canada}
\affil{Herzberg Astronomy \& Astrophysics Programs, National Research Council of Canada, 5071 West Saanich Road, Victoria, BC V9E 2E7, Canada}

\author{Ruobing Dong}
\affil{Department of Physics \& Astronomy, University of Victoria, Victoria, BC V8P 5C2, Canada}
\email{rbdong@uvic.ca}

\author{Takayuki Muto}
\affil{Division of Liberal Arts, Kogakuin University, 1-24-2 Nishi-Shinjuku, Shinjuku-ku, Tokyo 163-8677, Japan}

\author{Rebecca G. Martin}
\affil{Department of Physics \& Astronomy, University of Nevada, Las Vegas, 4505 South Maryland Parkway, Las Vegas, NV 89154, USA}

\author{Jeremy L. Smallwood}
\affil{Department of Physics \& Astronomy, University of Nevada, Las Vegas, 4505 South Maryland Parkway, Las Vegas, NV 89154, USA}

\author{Jun Hashimoto}
\affil{Astrobiology Center, National Institutes of Natural Sciences, 2-21-1 Osawa, Mitaka, Tokyo 181-8588, Japan}

\author{Hauyu Baobab Liu}
\affil{Academia Sinica Institute of Astronomy \& Astrophysics, No. 1, Sec. 4, Roosevelt Rd, Taipei 10617, Taiwan}

\author{Hideko Nomura}
\affil{Division of Science, National Astronomical Observatory of Japan, 2-21-1 Osawa, Mitaka, Tokyo 181-8588, Japan}
\affil{Department of Earth \& Planetary Sciences, Tokyo Institute of Technology, 2-12-1 Ookayama, Meguro, Tokyo 152-8551, Japan}

\author{Yasuhiro Hasegawa}
\affil{Jet Propulsion Laboratory, California Institute of Technology, Pasadena, CA 91109, USA}

\author{Michihiro Takami}
\affil{Academia Sinica Institute of Astronomy \& Astrophysics, No. 1, Sec. 4, Roosevelt Rd, Taipei 10617, Taiwan}

\author{Mihoko Konishi}
\affil{Faculty of Science \& Technology, Oita University, 700 Dannoharu, Oita 870-1192, Japan}

\author{Munetake Momose}
\affil{College of Science, Ibaraki University, 2-1-1 Bunkyo, Mito, Ibaraki 310-8512, Japan}

\author{Kazuhiro D. Kanagawa}
\affil{Research Center for the Early Universe, Graduate School of Science, University of Tokyo, Bunkyo, Tokyo 113-0033, Japan}

\author{Akimasa Kataoka}
\affil{Division of Science, National Astronomical Observatory of Japan, 2-21-1 Osawa, Mitaka, Tokyo 181-8588, Japan}

\author{Tomohiro Ono}
\affil{Department of Astrophysical Sciences, Princeton University, Princeton, NJ 08544, USA}
\affil{Department of Earth \& Space Science, Osaka University, Toyonaka, Osaka 560-0043, Japan}

\author{Michael L. Sitko}
\affil{Department of Physics, University of Cincinnati, Cincinnati, OH 45221, USA}
\affil{Space Science Institute, 475 Walnut Street, Suite 205, Boulder, CO 80301, USA}

\author{Sanemichi Z. Takahashi}
\affil{Division of Science, National Astronomical Observatory of Japan, 2-21-1 Osawa, Mitaka, Tokyo 181-8588, Japan}
\affil{Department of Applied Physics, Kogakuin University, 1-24-2 Nishi-Shinjuku, Shinjuku-ku, Tokyo 163-8677, Japan}

\author{Kengo Tomida}
\affil{Department of Earth \& Space Science, Osaka University, Toyonaka, Osaka 560-0043, Japan}
\affil{Astronomical Institute, Tohoku University, Sendai, Miyagi 980-8578, Japan}

\author{Takashi Tsukagoshi}
\affil{Division of Science, National Astronomical Observatory of Japan, 2-21-1 Osawa, Mitaka, Tokyo 181-8588, Japan}



\begin{abstract}

GW Ori is a hierarchical triple system with a rare circumtriple disk. We present Atacama Large Millimeter/submillimeter Array (ALMA) observations of 1.3 mm dust continuum and \textsuperscript{12}CO $J=2-1$ molecular gas emission of the disk. For the first time, we identify three dust rings in the GW Ori disk at $\sim$ 46, 188, and 338 AU, with estimated dust mass of 74, 168, and 245 Earth masses, respectively. To our knowledge, its outermost ring is the largest dust ring ever found in protoplanetary disks. We use visibility modelling of dust continuum to show that the disk has misaligned parts, and the innermost dust ring is eccentric. The disk misalignment is also suggested by the CO kinematics. We interpret these substructures as evidence of ongoing dynamical interactions between the triple stars and the circumtriple disk. 

\end{abstract}



\section{Introduction}

GW Ori is a hierarchical triple system \citep{berger2011} at a distance of 402 $\pm$ 10 parsec \citep{gaia2018}. Two of the stars (GW Ori AB) compose a spectroscopic binary with a separation of $\sim$ 1 AU \citep{mathieu1991}. A tertiary component (GW Ori C) was detected by near-infrared interferometry at a projected distance of $\sim$ 8 AU \citep{berger2011}. The stellar masses have been constrained to be $\sim$ 2.7, 1.7, and 0.9 $\msun$, respectively \citep{czekala2017}. The system harbours a rare circumtriple disk, with dust extending to $\sim$ 400 AU, and gas extending to $\sim$ 1300 AU \citep{fang2017}. Spectral energy distribution (SED) modelling indicates a gap in the disk at $25-55$ AU \citep{fang2014}.

Here we present high resolution ALMA observations in the disk around GW Ori at 1.3 mm dust continuum emission and \textsuperscript{12}CO $J=2-1$ emission, where we find new substructures of the disk that indicate ongoing disk-star interactions. We arrange the paper as follows: In Section \ref{sec:data}, we describe the setups of the ALMA observations and data reduction. In Section \ref{sec:obs}, we present the imaged results of dust continuum and \textsuperscript{12}CO $J=2-1$ observations. In Section \ref{sec:model}, we present results of dust continuum visibility modelling. In Section \ref{sec:discuss}, we discuss the possible origins of the observed substructures. In Section \ref{sec:sum}, we summarize our findings and raise some open questions.


\section{Observation and Data Reduction} \label{sec:data}

The observations were taken on December 10, 2017 (ID: 2017.1.00286.S). The disk was observed in Band 6 (1.3 mm) by 46 antennas, with baseline lengths ranging from 15 to 3321 meters. The total on source integration time was 1.6 hours. There were two 1.875-GHz-wide basebands centered at 217 and 233 GHz for continuum emission, and three basebands with 117 MHz bandwidths and 112 kHz resolution, centered at 230.518, 219.541 and 220.380 GHz to cover the \textsuperscript{12}CO, \textsuperscript{13}CO and C\textsuperscript{18}O $J=2-1$ lines. 

The data were calibrated by the pipeline calibration script provided by ALMA. We used the Common Astronomy Software Applications package \citep[\textsc{casa}; version 5.1.1-5;][]{mcmullin2007} to process the data. We adopted \textsc{casa} task \textsc{clean} to image the continuum map (Figure \ref{fig:obs}a), with the \textsc{uniform} weighting scheme and a 0".098 circular restoring beam. We performed phase self-calibration onto the image with a solution interval of 20 seconds. This resulted in an RMS noise level of $\sim$ 40 $\mu$Jy beam\textsuperscript{-1} and an enhanced peak signal-to-noise ratio (SNR) of $\sim$ 157, compared with $\sim$ 63 $\mu$Jy beam\textsuperscript{-1} and $\sim$ 90 before self-calibration, respectively. The integrated flux density of the disk (195 $\pm$ 20 mJy) is consistent with the result from previous ALMA observations \citep[202 $\pm$ 20 mJy;][]{czekala2017}.

The \textsuperscript{12}CO $J=2-1$ line data were obtained (after subtracting the continuum on the self-calibrated data) in the \textsc{briggs} weighting scheme with robust = 0.5, and velocity resolution 0.5 km s\textsuperscript{-1}. The resulting line cube has a beam size of 0".122 $\times$ 0".159 at the position angle -32.3\textsuperscript{$\circ$}. The noise level is $\sim$ 1.4 mJy beam\textsuperscript{-1} per channel and the peak signal-to-noise ratio is $\sim$ 83. Line emission was detected between 7.0 and 20.0 km s\textsuperscript{-1} with a central velocity of 13.5 km s\textsuperscript{-1}. The integrated flux is 60.6 Jy km s\textsuperscript{-1}, assuming a 2" radius. The intensity-weighted velocity map (a.k.a., the first-moment map) was constructed by calculating the intensity-weighted velocity with a threshold of three times the noise level. The averaged uncertainty of the twisted pattern in the first-moment map (i.e., the inset in Figure \ref{fig:obs}b) is $\sim$ 0.2 km s\textsuperscript{-1}, derived from error propagation theory, assuming the uncertainty of velocity due to the channel resolution is 0.25 km s\textsuperscript{-1}. The observations of the CO isotopologues C\textsuperscript{18}O and \textsuperscript{13}CO $J=2-1$ emission will be presented in future work.


\section{Observational Results} \label{sec:obs}

\subsection{Dust Continuum Emission}

Figure \ref{fig:obs}a shows the continuum map with spatial resolution of 0".098 ($\sim$ 39 AU). We identify three dust rings with different apparent shapes in the disk at $\sim$ 46, 188, and 338 AU (hereafter the inner, middle, and outer ring). The location of the inner ring coincides with the predicted cavity size from SED modelling \citep{fang2017}. The continuum flux densities of the inner, middle, and outer ring are 42 $\pm$ 4, 95 $\pm$ 10 and 58 $\pm$ 6 mJy, respectively. To our knowledge, the outer ring is the largest ever found in protoplanetary disks. 

The three rings harbor an enormous amount of solid material. We estimate the dust (solid) mass $M_{\text{dust}}$ of the rings with the equation provided in \cite{hildebrand1983} 
\begin{equation}
M_{\text{dust}}=\frac{F_{\nu}d^2}{B_{\nu}(T_{\text{dust}})\kappa_{\nu}}, 
\end{equation}
where $F_{\nu}$ is the continuum surface brightness at a sub-mm frequency $\nu$, \textit{d} is the distance from the observer to the source, $B_{\nu}(T_{\text{dust}})$ is the Planck function at the dust temperature $T_{\text{dust}}$, and $\kappa_{\nu}$ is the dust opacity. The dust temperature is estimated using a fitting function provided by \cite{dong2018b}
\begin{equation} \label{eq:tdust}
T_{\text{dust}}=30\left(\frac{L_{\star}}{38\,L_{\odot}}\right)^{1/4}\left(\frac{r}{100\,\textsc{AU}}\right)^{-1/2},
\end{equation}
where $L_{\star}$ is the total stellar luminosity, and \textit{r} is the location of the ring. The stellar luminosity modified by the distance provided by GAIA DR2 is 49.3 $\pm$ 7.4 $L_{\odot}$ \citep{calvet2004, gaia2018}. We assume a dust grain opacity of 10 cm\textsuperscript{2} g\textsuperscript{-1} at 1000 GHz with a power-law index of 1 \citep{beckwith1990}. We estimate the dust masses of the rings to be 74 $\pm$ 8, 168 $\pm$ 19, and 245 $\pm$ 28 $\mearth$, respectively, with the uncertainties incorporating the uncertainties in the surface brightness of the rings, source distance, stellar luminosity, and radial location of the rings.


\subsection{\textsuperscript{12}CO J=2-1 Emission} 

Figure \ref{fig:obs}b shows the first-moment map of \textsuperscript{12}CO $J=2-1$ emission (with the zeroth-moment map provided in Appendix \ref{app:mom0}). For regular Keplerian rotating disks, we expect a well-defined butterfly-like pattern in the first-moment map. However, we find a twisted pattern inside $\sim$ 0".2, which may result from a misalignment between the inner and outer parts of the disk \citep[i.e., having different inclinations and orientations;][]{rosenfeld2014}, as has been found in the disks around, e.g., HD 142527 \citep{casassus2015, marino2015} and HD 143006 \citep{perez2018, benisty2018}. 


\section{Modelling of Dust and Gas Emission} \label{sec:model}

The different apparent shapes of the rings could result from a few scenarios, such as coplanar rings with different eccentricities, circular rings with different inclinations, or rings with both different eccentricities and inclinations. Here we present evidence for disk misalignment and disk eccentricity found in modelling the dust and gas emission.

\subsection{Visibility Modelling of the Dust Continuum Emission} \label{mod_dust}

We fit the dust continuum map assuming that there are three dust rings in the disk with Gaussian radial profiles of surface brightness
\begin{equation}
F_i(r)=F_{0,i}e^{\frac{-(r-r_i)^2}{2\sigma_i^2}},
\end{equation}
where $F_i$ is the surface brightness as a function of the distance to the center \textit{r}, with \textit{i} = 1, 2, 3 denoting parameters for the inner, middle, and outer ring, respectively. $F_{0,i}$ is the peak surface brightness, $r_i$ is the radius of the ring (i.e., where the ring has the highest surface brightness), and $\sigma_i$ is the standard deviation. 
 
Initially, we assume all three rings are intrinsically circular when viewed face-on, and their different apparent shapes entirely originate from different inclinations. For each ring, we assume an independent set of peak surface brightness, center location, radius, width, inclination, and position angle as the model parameters. We call this combination of assumptions Model 1. 

After projecting the rings according to their position angles and inclinations, we calculate the synthetic visibility of the models using \textsc{galario} \citep{tazzari2018}, and launch MCMC parameter surveys to derive posterior distribution of model parameters using \textsc{emcee} \citep{dfm2013}. In the MCMC parameter surveys, the likelihood function $L$ is defined as 
\begin{equation} \begin{split}
\ln{L}=-\frac{1}{2}\sum^N_{j=1}m_j\times[(ReV_{\text{obs},j}-ReV_{\text{mod},j})^2&\\ +(ImV_{\text{obs},j}-ImV_{\text{mod},j})^2&] 
\end{split} \end{equation}
where $V_{\text{obs}}$ is the visibility data from ALMA observations, $V_{\text{mod}}$ is the synthetic model visibility, \textit{N} is the total number of visibility data points in $V_{\text{obs}}$, and $m_j$ is the weight of each visibility data point in $V_{\text{obs}}$. The prior function is set to guarantee the surface brightness, ring radius, and ring width do not go below zero, the position angle does not go beyond (-90, 90) degrees, and the inclination does not go beyond (0, 90) degrees. For each model, there are 144 chains spread in the hyperspace of parameters. Each chain has 15000 iterations including 10000 burn-in iterations. The results of the parameter surveys are listed in Table \ref{tab:dust}. 

The fitting result of Model 1, listed in Table \ref{tab:model1}a, suggests that the three rings have statistically different centers. Particularly, the center of the inner ring differs from the centers of the outer two by $\sim$ 20\% of the inner ring’s radius. This non-concentricity indicates non-zero intrinsic  eccentricities in the rings, particularly the inner ring (see Section \ref{sec:ecc}).

We explore the non-zero intrinsic eccentricity in the inner ring with two models. In both models, the outer two rings are intrinsically circular and concentric. Their center coincides with one of the two foci of the inner ring. In Model 2, that center is set free, while in Model 3 it is assumed to coincide with the stellar position provided by GAIA DR2 \citep{gaia2018}. In those two models, we introduce two more parameters for the intrinsic eccentricity and apoapsis angle of the inner ring. The position angle only indicates the direction to the ascending node on the axis along which the ring is inclined. The fitting results are listed in Table \ref{tab:model2}b and \ref{tab:model3}c, and the following calculations are based on the result of Model 3.

Figure \ref{fig:obs}c and \ref{fig:obs}e show the model image and the residual map of Model 3, respectively. The residual map is produced by subtracting model from data in the visibility plane, and then imaging the results in the same way used for the observations. We interpret the residuals as additional substructures on top of the ideal model \citep[e.g., a warp within the ring;][]{huang2020}.

All three models yield roughly consistent inclinations and position angles of each ring. However, we cannot determine the mutual inclinations between them (i.e., the angles between their angular momentum vectors) from dust emission modelling alone, due to the unknown direction of orbital motion.


\subsection{Kinematics Modelling of the \textsuperscript{12}CO $J=2-1$ Emission}

Following the prescription and parameter values used to fit low resolution CO isotopologue data of GW Ori \citep{fang2017}, we set up a gas surface density model using a power-law profile with an exponential tail
\begin{equation}
\Sigma(r)=\Sigma_c\left(\frac{r}{r_c}\right)^{-\gamma}e^{\left[-(r-r_c)^{2-\gamma}\right]},
\end{equation}
and the aspect ratio \textit{h/r} parametrized as 
\begin{equation} \label{eq:hr}
\frac{h}{r}=\left(\frac{h}{r}\right)_c\left(\frac{r}{r_c}\right)^{\psi},
\end{equation}
where $\Sigma_c$ and $(h/r)_c$ are corresponding values at the characteristic scaling radius $r_c$. The disk mass is taken as 0.12 $\msun$, corresponding to $\Sigma_c$ = 3 g cm$^{-2}$ for $r_c$ = 320 AU, with $\gamma$ = 1.0, $(h/r)_c$ = 0.18 and $\psi$ = 0.1. The dust surface density profile is set by assuming a gas-to-dust ratio of 100, and decreasing the dust surface density by a factor of 1000 inside the derived gap radii: inside 37 AU, from 56 to 153 AU, and from 221 to 269 AU. The \textsuperscript{12}CO channel maps are then computed and ray-traced by the physical-chemical modeling code \textsc{dali} \citep{bruderer2013}, which simultaneously solves the heating-cooling balance of the gas and chemistry to determine the gas temperature, molecular abundances and molecular excitation for a given density structure. 

\begin{figure*} 
\centering
\includegraphics[width = 0.8\textwidth]{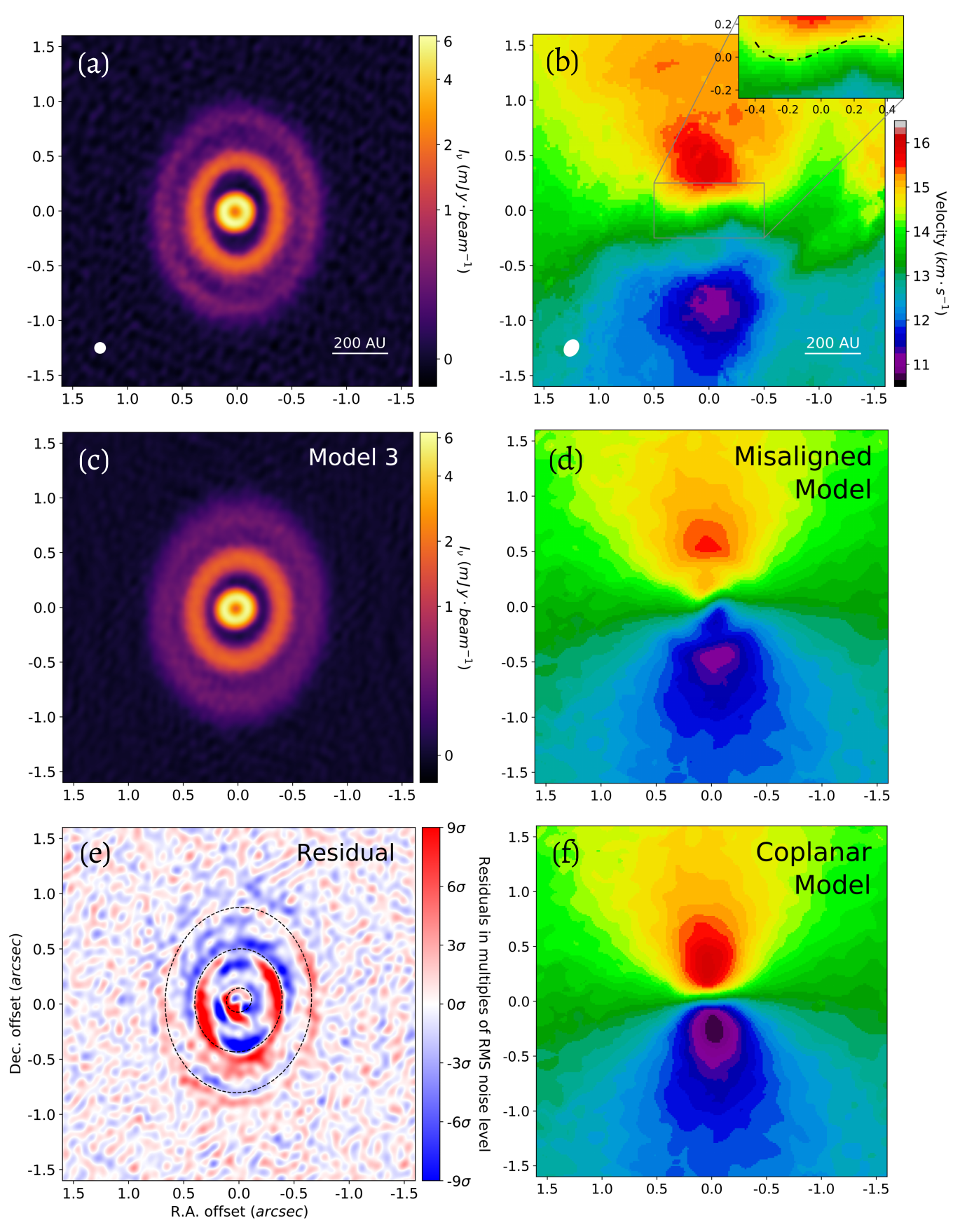}
\figcaption{
All panels are centered on the stellar position provided by GAIA DR2 (ICRS R.A. = $5^{\text{h}}29^{\text{m}}08^{\text{s}}.390$ and Dec. = 11\textsuperscript{$\circ$}52'12''.661).
(a): The ALMA self-calibrated dust continuum map performed with a 0".098 circular beam (bottom left corner; RMS noise level $\sigma \sim$ 40 $\mu$Jy beam\textsuperscript{-1}). A larger view of this panel is provided in Appendix \ref{app:large}.
(b): The ALMA \textsuperscript{12}CO J=2-1 first-moment map performed with a 0".122 × 0".159 beam with a position angle of -32.3\textsuperscript{$\circ$} (bottom left corner). The inset shows an 1" by 0".5 wide (40 by 20 AU) zoom, and the dot-dashed line highlights the shape of the twist. The averaged uncertainty in the inset region is $\sim$ 0.2 km s\textsuperscript{-1}. 
(c): Simulated ALMA continuum emission map of Model 3, produced in the same way as panel (a).
(d): The synthetic first-moment map of the misaligned disk model, applying the model parameters listed in Table \ref{tab:gas}. The color scheme is the same as panel (b).
(e): The residual map of Model 3. Dashed ellipses mark the fitted location of the three dust rings. The colorbar shows the residual magnitude in units of RMS noise level (1$\sigma$ = $\sim$ 40 $\mu$Jy beam\textsuperscript{-1} = $\sim 0.6\%$ of peak surface density).
(f): The synthetic first-moment map of the coplanar disk model, with an inclination of 37.9\textsuperscript{$\circ$} and a position angle of -5\textsuperscript{$\circ$} throughout the disk. The color scheme is the same as panel (b).
\label{fig:obs}}
\end{figure*}

Similar to \cite{walsh2017}, we model the misaligned disk with an inner cavity and three annuli each with its own inclination and position angle\footnote{\textsc{dali} is unable to vary inclination as a function of radius. The final channel map is constructed by concatenating three channel maps, each for one component, ray-traced at its inclination and position angle and cut out at the specified radius range listed in Table \ref{tab:gas}.}, as listed in Table \ref{tab:gas}. The channel map is run through the ALMA simulator using the settings of the ALMA observations. The resulting first-moment map is shown in Figure \ref{fig:obs}d. In Figure \ref{fig:obs}f we show the simulated first-moment map for another model as a comparison, in which the disk is coplanar with an inclination of 37.9\textsuperscript{$\circ$} and a position angle of -5\textsuperscript{$\circ$}.

The models show that the \textsuperscript{12}CO $J=2-1$ first-moment map in the ALMA observation cannot be reproduced by a coplanar disk. Instead, the misaligned disk model described in Table \ref{tab:gas} matches the observed first-moment map better, indicating the presence of misalignment in the GW Ori disk.


\section{Discussions} \label{sec:discuss}

Several disks have been observed to have non-zero eccentricity and/or misalignment (e.g., MWC 758, \citealt{dong2018a}; HD 142527, \citealt{casassus2015, marino2015}; and HD 143006, \citealt{perez2018, benisty2018}). Unlike most of them, in which the origin is uncertain, the GW Ori system provides strong and direct link between substructures and star-disk gravitational interactions. Therefore it offers a unique laboratory to probe three-dimensional star-disk interactions. In this section, we discuss the possible origins of the observed substructures due to star-disk interactions. 


\subsection{Disk Eccentricity} \label{sec:ecc}

The A-B binary and the C component can be dynamically viewed as an AB-C binary. The eccentricity of the circumbinary disk may increase through resonant interactions with the binary \citep{papaloizou2001}. In the case of no binary-disk misalignment, the binary’s perturbing gravitational potential on the midplane of the disk is given by \cite{lubow1991a}. The coupling of this perturbing potential with the imposed eccentricity of the disk excites density waves at the 1:3 outer eccentric Lindblad Resonance, which lead to angular momentum removal in the inner parts of the disk. As no energy is removed along with the angular momentum in this process, the disk orbit cannot remain circular \citep{papaloizou2001}. In the case of GW Ori, the inner dust ring is the most susceptible to this effect, which could explain why its center in Model 1 is more deviated from those of the other two rings. 


\subsection{Binary-disk Misalignment} \label{sec:deg}

\begin{figure*} 
\includegraphics[width = \textwidth]{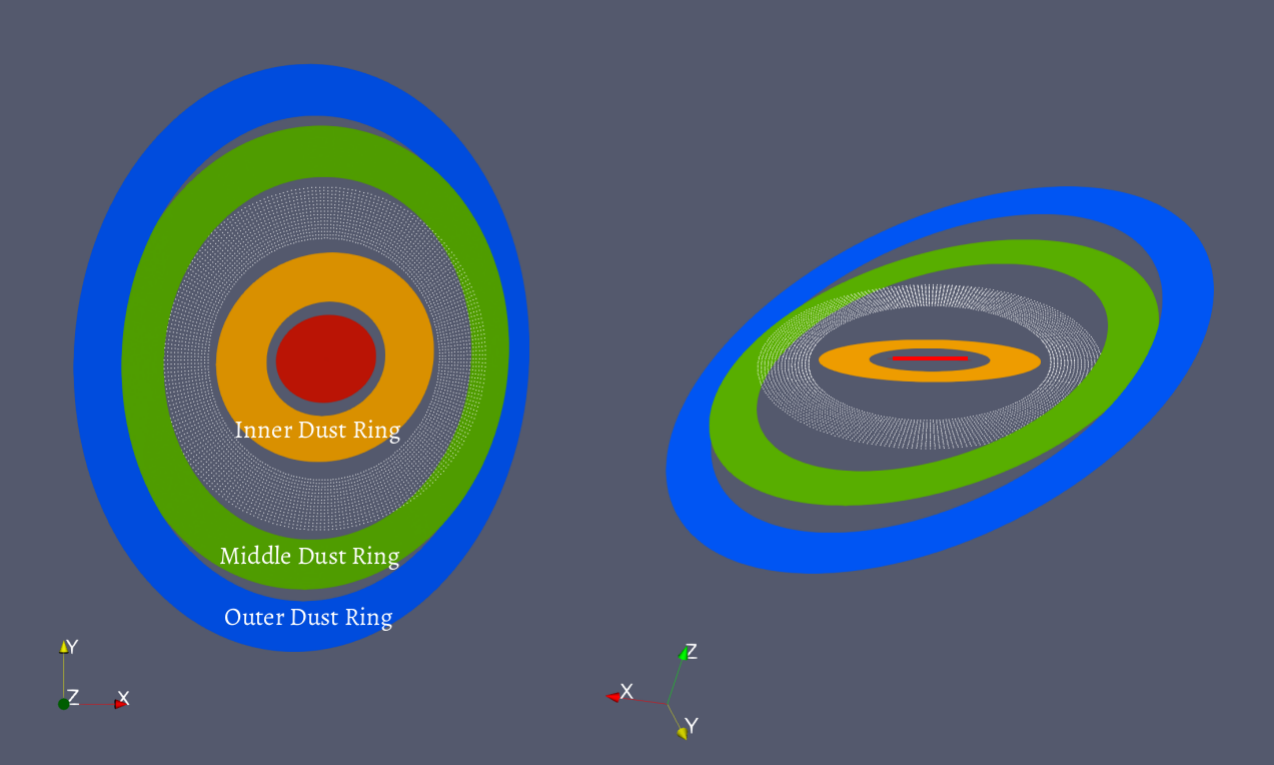}
\figcaption{A schematic diagram showing the proposed geometry of the system. Orbital planes of the AB-C binary (red), the inner dust ring (orange), the gap between the inner and middle dust rings (white dots), the middle dust ring (green), and the outer dust ring (blue) are marked inside out. The left panel is a sky-projected view, and in the right panel the binary is edge-on. The size of the disk components are not to-scale. The orientation axes are shown at the bottom left corner of each panel, with \textit{x}-axis being antiparallel to the R.A. direction, \textit{y}-axis being parallel to the Dec. direction, and \textit{z}-axis pointing at the observer. 
\label{fig:scheme}} 
\end{figure*}

Our dust and gas observations alone cannot break the degeneracy in the mutual inclination between different parts in the disk due to the unknown on-sky projected orbital direction of the disk. Previous studies indicate that the on-sky projected gas motion is likely to be clockwise \citep{czekala2017}, same as the orbital motion of GW Ori C given by astrometric observations \citep{berger2011}. Given the inclination and longitude of ascending node of the AB-C binary orbit being 150 $\pm$ 7 and 282 $\pm$ 9 degrees \citep{czekala2017}, we assume that the entire disk has the same clockwise on-sky projected orbital direction. Following \citet{fekel1981}, we find out that the binary-disk misalignments at 46 AU (the inner ring), ~100 AU (a gap), 188 AU (the middle ring), and 338 AU (outer ring) are 11 $\pm$ 6, $\sim$28\footnote{the manual fitting of gas model cannot provide any uncertainties for the gap between the inner and middle ring}, 35 $\pm$ 5, and 40 $\pm$ 5 degrees, respectively. A schematic diagram of our disk model is displayed in Figure \ref{fig:scheme}. Therefore, the inner ring and the AB-C binary plane are close to being coplanar, and there is a monotonic trend of binary-disk misalignment from $\sim$ 10\textsuperscript{$\circ$} at $\sim$ 50 AU to $\sim$ 40\textsuperscript{$\circ$} at $\sim$ 340 AU, consistent with the expected outcome of the disk misalignment (see Section \ref{sec:warp}).

Several mechanisms could produce an initial binary-disk misalignment, such as turbulence in the star-forming gas clouds \citep{bate2012}, binary formation in the gas cloud whose physical axes are misaligned to the rotation axis \citep{bonnell1992}, and accretion of cloud materials with misaligned angular momentum with respect to the binary after the binary formation \citep{bate2018}.


\subsection{Misalignment Within the Disk} \label{sec:warp}

A test particle orbiting a binary on a misaligned orbit undergoes nodal precession due to gravitational perturbations from the binary \citep{nixon2011, facchini2018}. For a protoplanetary disk in the bending-wave regime \citep[i.e., where the aspect ratio is higher than the $\alpha$-prescription of viscosity;][]{shakura1973}, disk parts at different radii shall undergo global precession like a rigid body with possibly a small warp \citep{smallwood2019}. Therefore the timescale of radial communication of disk materials (i.e., for pressure-induced bending waves propagating at half of the sound speed) and the timescale of global precession determine whether the disk can develop a misalignment inside. 

Assuming an inner radius at 32 AU \citep[$3-4$ times the AB-C binary semi-major axis;][]{czekala2017, kraus2020} and an outer radius at 1300 AU (size of the gas disk, \citealt{fang2017}), the global precession time-scale of the entire disk is $\sim$ 0.83 Myr, and the radial communication time-scale is $\sim$ 0.06 Myr (see Appendix \ref{app:tc} and \ref{app:tn} for detailed calculations). The radial communication is able to prevent the disk from breaking or developing a significant warp, and we would not expect the observed large deviations in the inclination and position angle between the inner and middle ring. Therefore, we propose that the gap between the inner and middle ring is deep enough to break the disk into two parts (hereafter the inner and outer disk), undergoing nodal precession independently, due to another mechanism. 

Due to the viscous dissipation, the precessing disk is torqued toward either polar alignment (i.e., the binary-disk misalignment becomes 90\textsuperscript{$\circ$}; \citealt{martin2017, zanazzi2018}), or coplanar alignment/counter-alignment. The minimum critical initial binary-disk misalignment for which a disc moves towards polar alignment is $\sim 63^{\circ}$ in the limit of zero disk mass. Since a higher disk mass will lead to a larger critical angle \citep{martin2019}, the GW Ori disk is most likely moving towards coplanar alignment. As we propose that the disk breaks into two parts undergoing global precession independently, they are also aligning to the binary independently on different time-scales. 

Assuming the radial communication is blocked at 60 AU, we estimate the alignment time-scales to be $\sim$ 1 Myr for the inner disk and above 100 Myr for the outer disk (see Appendix \ref{app:ta}). This is consistent with the observed significantly smaller inclination of the inner ring with respect to the binary than those of the outer rings. The latter are likely inherited from birth and have not evolved much given the system age. 

If the radial communication is also blocked at the gap between the middle and outer ring (e.g., $\sim$ 250 AU), the nodal precession time-scales for the middle and outer ring would be $\sim$ 0.6 Myr and $\sim$ 120 Myr, respectively, and the two rings are likely to develop significantly different position angles. Thus we propose that the gap between the middle and outer ring does not cut off the radial communication, and the two rings precess roughly as a rigid body with only a small warp between them.


\subsection{Hydrodynamic Simulations} \label{sec:sim}

\begin{figure*} 
\includegraphics[width = 0.5\textwidth]{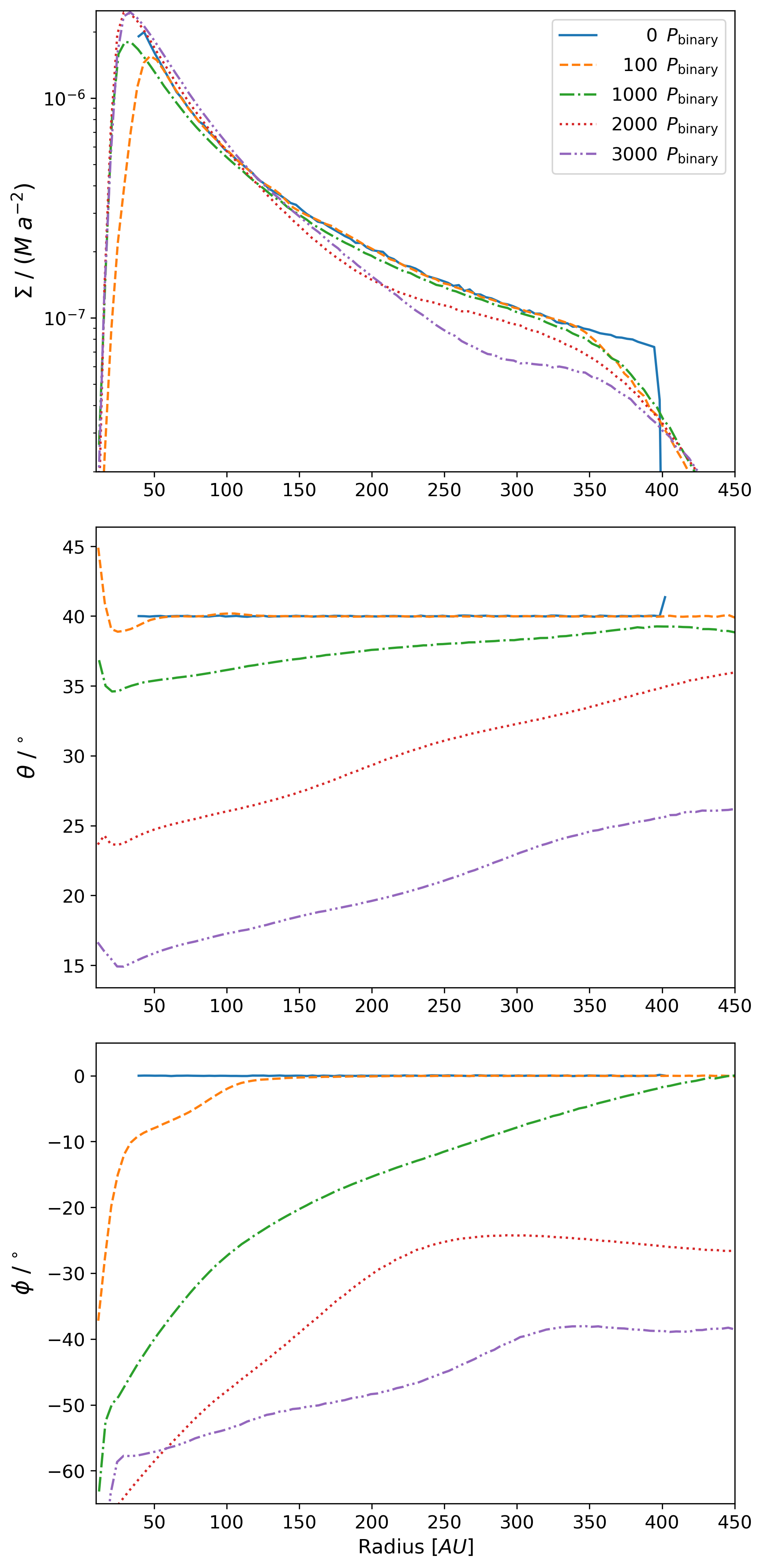}
\figcaption{The result of the SPH simulation. \textit{Upper panel:} Radial profile of the surface density of the disk. \textit{M} and \textit{a} are the total mass and separation of the AB-C binary in code units, respectively. \textit{Middle panel:} Radial profile of the binary-disk misalignment. \textit{Lower panel:} Radial profile of the longitude of the ascending node of the disk, measured from the binary’s orbital plane.
\label{fig:sim}} 
\end{figure*}

The analytical results suggest that the radial communication is able to prevent the binary from breaking the disk \citep[e.g.][]{nixon2013}. As a result, we propose a break at $\sim$ 60 AU that is due to other mechanisms in order to explain the observed structures. We carry out a demonstrative smoothed particle hydrodynamic (SPH) simulation with the \textsc{phantom} code \citep{lodato2010, price2010, price2018} to test the non-breaking hypothesis in the non-linear regime.  The results are shown in Figure \ref{fig:sim}.

We model the triple star system as the outer binary in order to speed up the simulation. The simulation consists of 10\textsuperscript{6} equal mass Lagrangian SPH particles initially distributed from $r_{\text{in}}$ = 40 AU to $r_{\text{out}}$ = 400 AU. The initial truncation radius of the disk does not affect the simulation significantly, since the material moves inwards quickly due to the short local viscous timescale. A smaller initial outer truncation radius $r_{\text{out}}$ than what is observed is chosen in order to better resolve the disk. The binary begins at apastron with $e_{\text{b}}$ = 0.22 and $a_{\text{b}}$ = 9.2 AU \citep{czekala2017}. The accretion radius of each binary component is 4 AU. Particles within this radius are accreted, and their mass and angular momentum are added to the star. We ignore the effect of self-gravity since it has no effect on the nodal precession rate of flat circumbinary disks. 

The initial surface density profile is taken by
\begin{equation}
\Sigma(r)=\Sigma_0\,(r/r_0)^{-3/2},
\end{equation}
where $\Sigma_0$ is the density normalization at $r_0$ = 40 AU, corresponding to a total disk mass of 0.1 $\msun$. We take a locally isothermal disk with a constant aspect ratio $h/r=0.05$, where $h$ is the scale-height. The \cite{shakura1973} $\alpha$ parameter varies in the range $0.008 - 0.013$ over the disk. The SPH artificial viscosity  $\alpha_{\rm AV} = 0.31$ mimics a disk with  
\begin{equation}
\alpha \approx \frac{\alpha_{\textsc{av}}}{10}\frac{\langle H \rangle}{h}
\end{equation}
 \cite{lodato2010}, where $\langle H \rangle$ is the mean smoothing length on particles in a cylindrical ring at a given radius and we take $\beta_{\rm AV}=2$. The disk is resolved with average smoothing length per scale height of $0.32$.

The evolution of surface density, binary-disk misalignment, and longitude of the ascending node at different radii suggest that the disk does not show any sign of breaking in 3000 binary’s orbital periods ($\sim$ 0.04 Myr), which is sufficiently long to tell if the disk would break or not since the radial communication time-scale in the simulation is $\sim$ 0.01 Myr. Instead, the disk presents a global warp. The warp is not taken into account in the analytic estimates. The small outer truncation radius in the simulation leads to a faster precession time-scale than that predicted by the analytic model, comparable to the radial communication time-scale. The simulations suggests that unless the disk is very cool (i.e., low aspect ratio) and in the viscous regime, some other mechanism, e.g., a companion, is needed to break the disk at the gap between the inner and middle ring. This mechanism may also be responsible for producing the observed misalignment in the disk.

A disk with a lower aspect ratio or a higher $\alpha$ value, such that it falls into the viscous regime ($h/r < \alpha$), may break due to the binary’s torque \citep{nixon2013}. However, observations have suggested lower $\alpha$ values than our adaption here \citep[e.g., $\alpha\lesssim10^{-3}$,][]{flaherty2017}. A lower viscosity leads to a larger binary truncation radius \citep{artymowicz1994}, and a longer global precession time-scale (Equation \ref{eq:global}). Therefore we expect even less warping (and no break) in the GW Ori disk than in our simulation.


\section{Conclusions} \label{sec:sum}

We present ALMA 1.3 mm dust continuum observation and \textsuperscript{12}CO $J=2-1$ emission of the circumtriple disk around GW Ori. Our main conclusions are the following: 

\begin{enumerate}

\item For the first time, we identify three dust rings in the GW Ori disk at $\sim$ 46, 188, and 338 AU, with their estimated dust mass being $\sim$ 74, 168, and 245 $\mearth$, respectively. The three dust rings have enough solids to make many cores of giant planets \citep[$\sim$ 10 $\mearth$;][]{pollack1996}.

\item We built three models under various assumptions to fit the dust continuum observations using MCMC fitting. Our results (Table \ref{tab:dust}) suggest that the inner ring has an eccentricity of  $\sim$ 0.2, and the three rings have statistically different on-sky projected inclinations. The inner, middle, and outer ring are likely misaligned by $\sim$ 11, 35, and 40 degrees to the orbital plane of the GW Ori AB-C binary system, respectively.

\item A twisted pattern is identified in the first-moment map, suggesting the presence of a warp in the disk, consistent with what we have found in the dust continuum emission. 

\item Using analytical analysis and hydrodynamic simulations, we find that the torque from the GW Ori triple stars alone cannot explain the observed large misalignment between the inner and middle dust rings. The disk would not break due to the torque, and a continuous disk is unlikely to show the observed large misalignment. Therefore this hints at some other mechanism that breaks the disk and prevents radial communication of bending waves between the inner and middle ring.

\end{enumerate}

There are still open questions associated with the system. For example, are there any companions in the disk? Dust rings and gaps have been shown to be common in protoplanetary disks \citep{long2018, andrews2018, huang2018, vandermarel2019}, and one of the most exciting hypotheses is that they are produced by embedded companions ranging from stellar-mass all the way to super-Earths \citep{artymowicz1994, dong2015, zhang2018}. Specifically, a companion may be opening the gap between the inner and middle ring and break the disk there. Companions at hundreds of AU from their host stars have been found before \citep[e.g., HD 106906 b;][]{bailey1994}. But how they form, i.e., forming in situ or at closer distances, or followed by scattering or migration to the outer regions, is unclear. If GW Ori’s dust rings are in the process of forming companions, there will be circumtriple companions, which have not been found before \citep[excluding quadruple systems;][]{busetti2018}. The system will offer direct clues on the formation of distant companions. 

\bigbreak

We thank Sean Andrews, Myriam Benisty, John Carpenter, Ian Czekala, Sheng-Yuan Liu, Feng Long, Diego Muñoz, Henry Ngo, Laura Pérez, John Zanazzi, and Zhaohuan Zhu for discussions. We also thank the anonymous referee for constructive suggestions that largely improved the quality of the paper. J.B. thanks Belaid Moa for helps on the numerical implementation. This paper makes use of the following ALMA data: ADS/JAO.ALMA\#2017.1.00286.S. ALMA is a partnership of ESO (representing its member states), NSF (USA) and NINS (Japan), together with NRC (Canada), MOST and ASIAA (Taiwan), and KASI (Republic of Korea), in cooperation with the Republic of Chile. The Joint ALMA Observatory is operated by ESO, AUI/NRAO and NAOJ. The National Radio Astronomy Observatory is a facility of the National Science Foundation operated under cooperative agreement by Associated Universities, Inc. Numerical calculations are performed on the clusters provided by ComputeCanada. This work is in part supported by JSPS KAKENHI Grant Numbers 19K03932, 18H05441 and 17H01103 and NAOJ ALMA Scientific Research Grant Number 2016-02A. Financial support is provided by the Natural Sciences and Engineering Research Council of Canada through a Discovery Grant awarded to R.D.. N.v.d.M. acknowledges support from the Banting Postdoctoral Fellowships program, administered by the Government of Canada. R.G.M. acknowledges support from NASA through grant NNX17AB96G. H.B.L. is supported by the Ministry of Science and Technology (MoST) of Taiwan (Grant Nos. 108-2112-M-001-002-MY3). Y.H. is supported by the Jet Propulsion Laboratory, California Institute of Technology, under a contract with the National Aeronautics and Space Administration. K.T is supported by JSPS KAKENHI 16H05998 and 18H05440, and NAOJ ALMA Scientific Research Grant 2017-05A.


\clearpage
\bibliography{jiaqing}

\clearpage

\clearpage
\begin{table}[t]
\centering
\caption{The Complete MCMC Result of Dust Continuum Visibility Modelling}
\label{tab:dust}

\medskip
(a) Model 1 \label{tab:model1} \\[3pt]

\begin{tabular*}{\textwidth}{@{\extracolsep{\fill}}crrr}
\toprule
    & \mc{Inner Ring} & \mc{Middle Ring} & \mc{Outer Ring} \\
\midrule
R.A Offset $[arcsecond]$
    & $ 1.89\times10^{-2}\;^{+9.47\times10^{-5}}_{-9.49\times10^{-5}}$
    & $-2.44\times10^{-3}\;^{+1.82\times10^{-4}}_{-2.08\times10^{-4}}$ 
    & $-4.24\times10^{-3}\;^{+4.64\times10^{-4}}_{-3.91\times10^{-4}}$ \\ \\[-1em]
Dec. Offset $[arcsecond]$
    & $-1.32\times10^{-2}\;^{+1.09\times10^{-4}}_{-1.01\times10^{-4}}$
    & $-2.26\times10^{-2}\;^{+2.22\times10^{-4}}_{-2.06\times10^{-4}}$
    & $-1.21\times10^{-2}\;^{+6.57\times10^{-4}}_{-4.67\times10^{-4}}$ \\ \\[-1em]
Ring Radius $[arcsecond]$
    & $ 1.15\times10^{-1}\;^{+1.46\times10^{-4}}_{-1.48\times10^{-4}}$
    & $ 4.68\times10^{-1}\;^{+3.07\times10^{-4}}_{-2.92\times10^{-4}}$
    & $ 8.40\times10^{-1}\;^{+1.03\times10^{-3}}_{-1.15\times10^{-3}}$ \\ \\[-1em]
Ring Width $[arcsecond]$
    & $ 4.97\times10^{-2}\;^{+3.16\times10^{-4}}_{-6.24\times10^{-4}}$
    & $ 1.74\times10^{-1}\;^{+6.69\times10^{-4}}_{-7.56\times10^{-4}}$
    & $ 3.32\times10^{-1}\;^{+1.84\times10^{-3}}_{-2.78\times10^{-3}}$ \\ \\[-1em]
Surface Brightness $[Jy/pixel]$  
    & $ 2.49\times10^{-4}\;^{+1.80\times10^{-6}}_{-1.56\times10^{-6}}$ 
    & $ 3.96\times10^{-5}\;^{+8.67\times10^{-8}}_{-1.69\times10^{-7}}$ 
    & $ 1.13\times10^{-5}\;^{+2.77\times10^{-8}}_{-5.52\times10^{-7}}$ \\ \\[-1em]
Inclination $[degree]$
    & $ 22.24\;^{+0.23}_{-0.31}$
    & $ 32.62\;^{+0.07}_{-0.11}$
    & $ 37.93\;^{+0.09}_{-0.08}$ \\ \\[-1em]
Position Angle $[degree]$
    & $-60.75\;^{+1.06}_{-0.56}$
    & $- 7.43\;^{+0.19}_{-0.12}$
    & $- 3.57\;^{+0.15}_{-0.12}$ \\ \\[-1em]
\bottomrule
\end{tabular*}

\medskip
(b) Model 2 \label{tab:model2} \\[3pt]

\begin{tabular*}{\textwidth}{@{\extracolsep{\fill}}crrr}
\toprule
    & \mc{Inner Ring} & \mc{Middle Ring} & \mc{Outer Ring} \\
\midrule
R.A Offset $[arcsecond]$
    &  & $ 1.77\times10^{-2}\;^{+1.69\times10^{-4}}_{-1.61\times10^{-4}}$ & \\ \\[-1em]
Dec. Offset $[arcsecond]$
    &  & $-2.22\times10^{-2}\;^{+1.95\times10^{-4}}_{-1.99\times10^{-4}}$ & \\ \\[-1.2em]
\midrule
Ring Radius $[arcsecond]$
    & $ 1.17\times10^{-1}\;^{+1.36\times10^{-4}}_{-1.35\times10^{-4}}$
    & $ 4.68\times10^{-1}\;^{+2.88\times10^{-4}}_{-2.71\times10^{-4}}$
    & $ 8.40\times10^{-1}\;^{+9.48\times10^{-4}}_{-9.53\times10^{-4}}$ \\ \\[-1em]
Ring Width $[arcsecond]$
    & $ 4.97\times10^{-2}\;^{+3.16\times10^{-4}}_{-3.91\times10^{-4}}$
    & $ 1.74\times10^{-1}\;^{+6.66\times10^{-4}}_{-6.03\times10^{-4}}$
    & $ 3.30\times10^{-1}\;^{+1.68\times10^{-3}}_{-1.96\times10^{-3}}$ \\ \\[-1em]
Surface Brightness $[Jy/pixel]$  
    & $ 2.51\times10^{-4}\;^{+1.53\times10^{-6}}_{-1.55\times10^{-6}}$ 
    & $ 3.96\times10^{-5}\;^{+8.20\times10^{-8}}_{-1.01\times10^{-7}}$ 
    & $ 1.13\times10^{-5}\;^{+2.65\times10^{-8}}_{-3.37\times10^{-8}}$ \\ \\[-1em]
Inclination $[degree]$
    & $ 23.15\;^{+0.22}_{-0.23}$
    & $ 32.64\;^{+0.07}_{-0.07}$
    & $ 37.91\;^{+0.08}_{-0.07}$ \\ \\[-1em]
Position Angle $[degree]$
    & $-55.67\;^{+0.61}_{-0.50}$
    & $- 7.44\;^{+0.14}_{-0.12}$
    & $- 3.60\;^{+0.13}_{-0.11}$ \\ \\[-1em]
Apoapsis Angle $[degree]$
    & $ 65.04\;^{+0.50}_{-0.49}$ & --- & --- \\ \\[-1em]
Eccentricity 
    & $ 0.21\;^{+1.75\times10^{-3}}_{-1.43\times10^{-3}}$ & --- & --- \\ \\[-1em]
\bottomrule
\end{tabular*}

\medskip
(c) Model 3 \label{tab:model3} \\[3pt]

\begin{tabular*}{\textwidth}{@{\extracolsep{\fill}}crrr}
\toprule
    & \mc{Inner Ring} & \mc{Middle Ring} & \mc{Outer Ring} \\
\midrule
Ring Radius $[arcsecond]$
    & $ 1.16\times10^{-1}\;^{+1.20\times10^{-4}}_{-1.49\times10^{-4}}$
    & $ 4.68\times10^{-1}\;^{+2.72\times10^{-4}}_{-2.98\times10^{-4}}$
    & $ 8.37\times10^{-1}\;^{+8.95\times10^{-4}}_{-1.06\times10^{-3}}$ \\ \\[-1em]
Ring Width $[arcsecond]$
    & $ 4.99\times10^{-2}\;^{+3.06\times10^{-4}}_{-3.86\times10^{-4}}$
    & $ 1.73\times10^{-1}\;^{+6.33\times10^{-4}}_{-5.98\times10^{-4}}$
    & $ 3.39\times10^{-1}\;^{+1.90\times10^{-3}}_{-1.85\times10^{-3}}$ \\ \\[-1em]
Surface Brightness $[Jy/pixel]$  
    & $ 2.47\times10^{-4}\;^{+1.64\times10^{-6}}_{-1.42\times10^{-6}}$ 
    & $ 3.94\times10^{-5}\;^{+8.55\times10^{-8}}_{-1.12\times10^{-7}}$ 
    & $ 1.13\times10^{-5}\;^{+2.63\times10^{-8}}_{-3.33\times10^{-8}}$ \\ \\[-1em]
Inclination $[degree]$
    & $ 20.63\;^{+0.23}_{-0.29}$
    & $ 32.86\;^{+0.06}_{-0.07}$
    & $ 37.96\;^{+0.09}_{-0.08}$ \\ \\[-1em]
Position Angle $[degree]$
    & $-60.37\;^{+0.81}_{-0.65}$
    & $- 7.26\;^{+0.13}_{-0.13}$
    & $- 3.49\;^{+0.13}_{-0.12}$ \\ \\[-1em]
Apoapsis Angle $[degree]$
    & $121.39\;^{+0.26}_{-0.25}$ & --- & --- \\ \\[-1em]
Eccentricity 
    & $ 0.19\;^{+8.55\times10^{-4}}_{-6.46\times10^{-4}}$ & --- & --- \\ \\[-1em]
\bottomrule
\end{tabular*}
\end{table}

\tablecomments{The radius of each ring is the location of the peak in our model in Section \ref{sec:model}, and the width is the full width at half maximum (FWHM) of the profile. The center offsets for Model 1 and Model 2 are relative to the center in Model 3, which is the location of GW Ori provided by GAIA DR2 (ICRS R.A. = $5^{\text{h}}29^{\text{m}}08^{\text{s}}.390$ and Dec. = 11\textsuperscript{$\circ$}52'12''.661). The position angles and apoapsis angles are measured East of North. The inclination is defined in the range from 0\textsuperscript{$\circ$} to 90\textsuperscript{$\circ$}, with 0\textsuperscript{$\circ$} denoting face-on. The pixel size in the unit of surface brightness is determined internally by \textsc{galario}.}


\clearpage
\begin{table}[t]
\centering
\caption{The Parameters Used in the Gas Kinematics Modelling}
\label{tab:gas}

\medskip
\begin{tabular*}{0.6\textwidth}{@{\extracolsep{\fill}}cccc}
\toprule
\multirow{2}{*}{Inner Ring} & \multirow{2}{*}{Outer Ring} & \mc{Longitude of the} & \multirow{2}{*}{Inclination} \\
                            &                             & \mc{Ascending Node}   &                              \\
\mc{[$AU$]}                 & \mc{[$AU$]}                 & \mc{[$degree$]}       & \mc{[$degree$]}              \\
\midrule
0   & 32   & \multicolumn{2}{c}{No Emission} \\ \\[-1em]
32  & 48   & -60 & 22.3                      \\ \\[-1em]
48  & 153  & -10 & 17.3                      \\ \\[-1em]
153 & 1000 & -5  & 37.9                      \\ \\[-1em]
\bottomrule
\end{tabular*}
\end{table}

\tablecomments{The annuli are concentric with the center located at the stellar position provided by GAIA DR2 (ICRS R.A. = $5^{\text{h}}29^{\text{m}}08^{\text{s}}.390$ and Dec. = 11\textsuperscript{$\circ$}52'12''.661). The longitude of the ascending node is measured East of North, and the inclination is defined in the range from 0\textsuperscript{$\circ$} to 90\textsuperscript{$\circ$} with 0\textsuperscript{$\circ$} meaning face-on. The disk model is composed of an empty inner cavity and three annuli, inside out. The inner annulus (from 32 to 48 AU) is for the inner dust ring. The outer annulus (from 153 to 1000 AU) is for the middle and outer dust rings. The middle annulus (from 48 to 153 AU) is for the gap in between.}

\clearpage
\appendix


\section{ALMA \textsuperscript{12}CO $J=2-1$ Zeroth-moment Map} \label{app:mom0}

\begin{figure*}[h]
\begin{center}
\includegraphics[width = 0.75\textwidth]{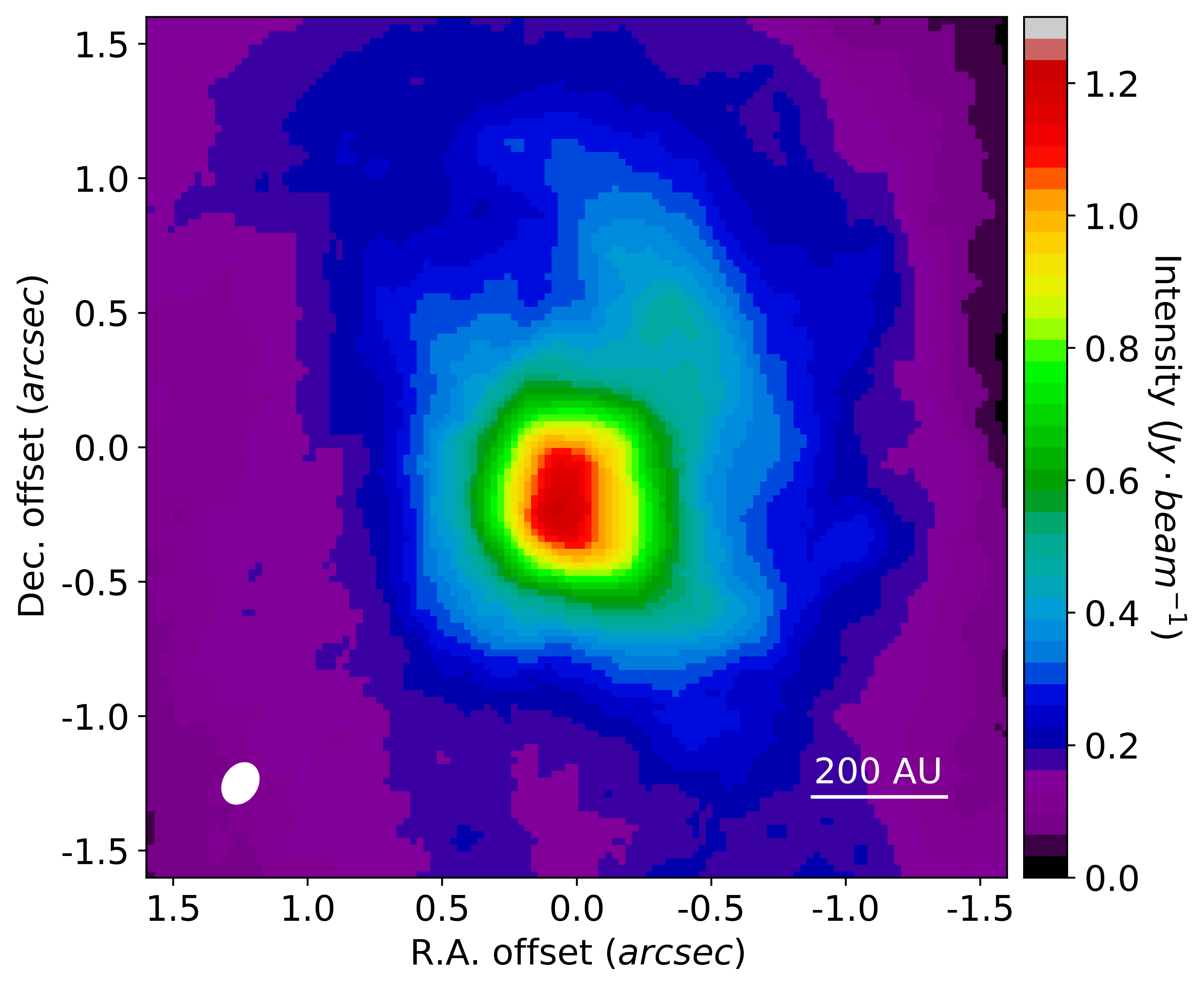}
\figcaption{The ALMA zeroth-moment map of \textsuperscript{12}CO $J=2-1$ emission.
\label{fig:mom0}} 
\end{center}
\end{figure*}

Figure \ref{fig:mom0} shows the ALMA zeroth-moment map of \textsuperscript{12}CO $J=2-1$ emission. The spiral-like structure at $\sim 0.75''$ to the northwest is likely due to cloud contamination, as reported by previous studies \citep{czekala2017, fang2017}.

\clearpage


\section{A Larger View of Figure 1a} \label{app:large}

\begin{figure*}[h]
\begin{center}
\includegraphics[width = \textwidth]{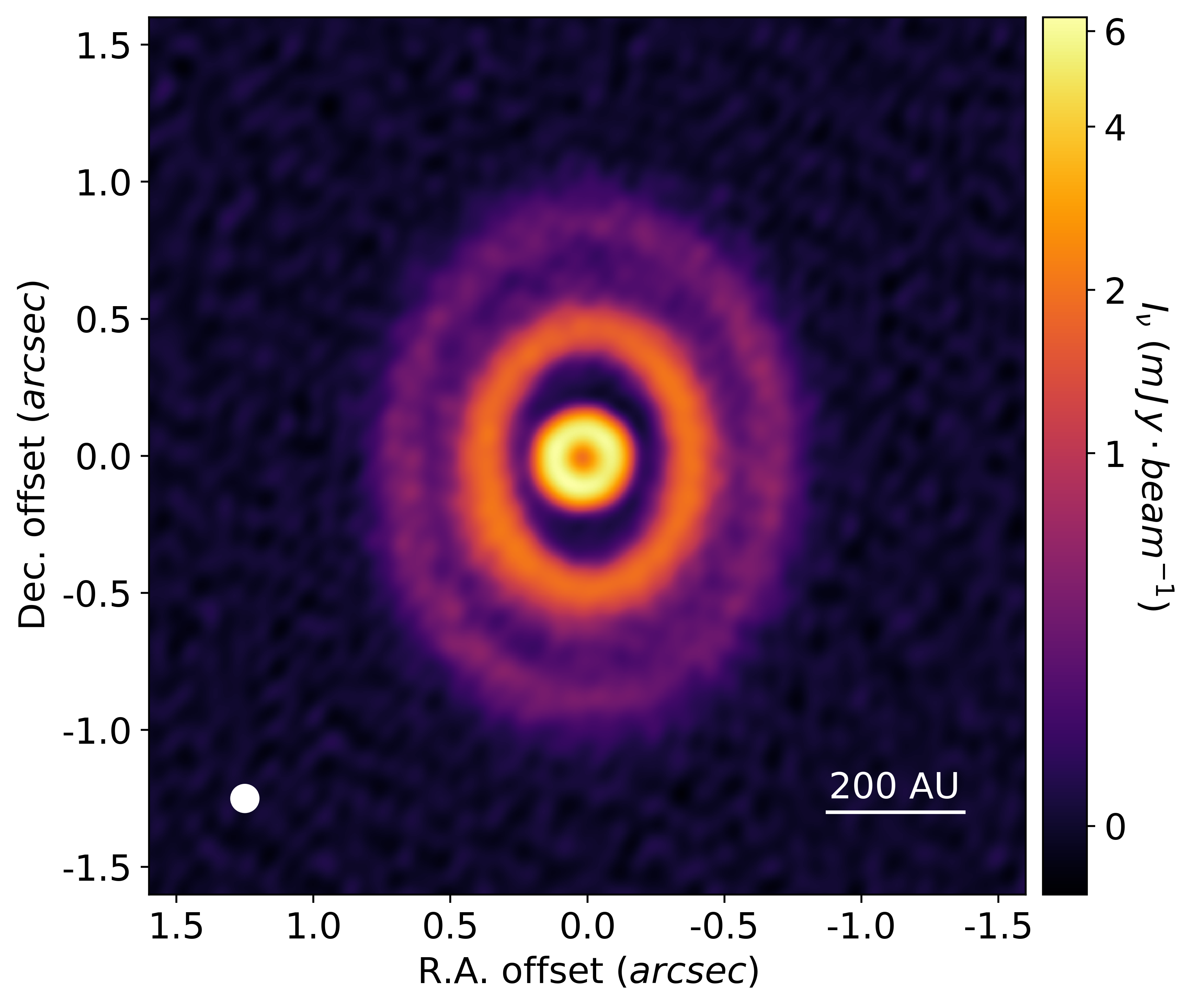}
\figcaption{A larger view of Figure \ref{fig:obs}a.} 
\end{center}
\end{figure*}


\section{the uv-plot and posterior distribution of dust modelling}

\begin{sidewaysfigure}[h]
\includegraphics[width = \textwidth]{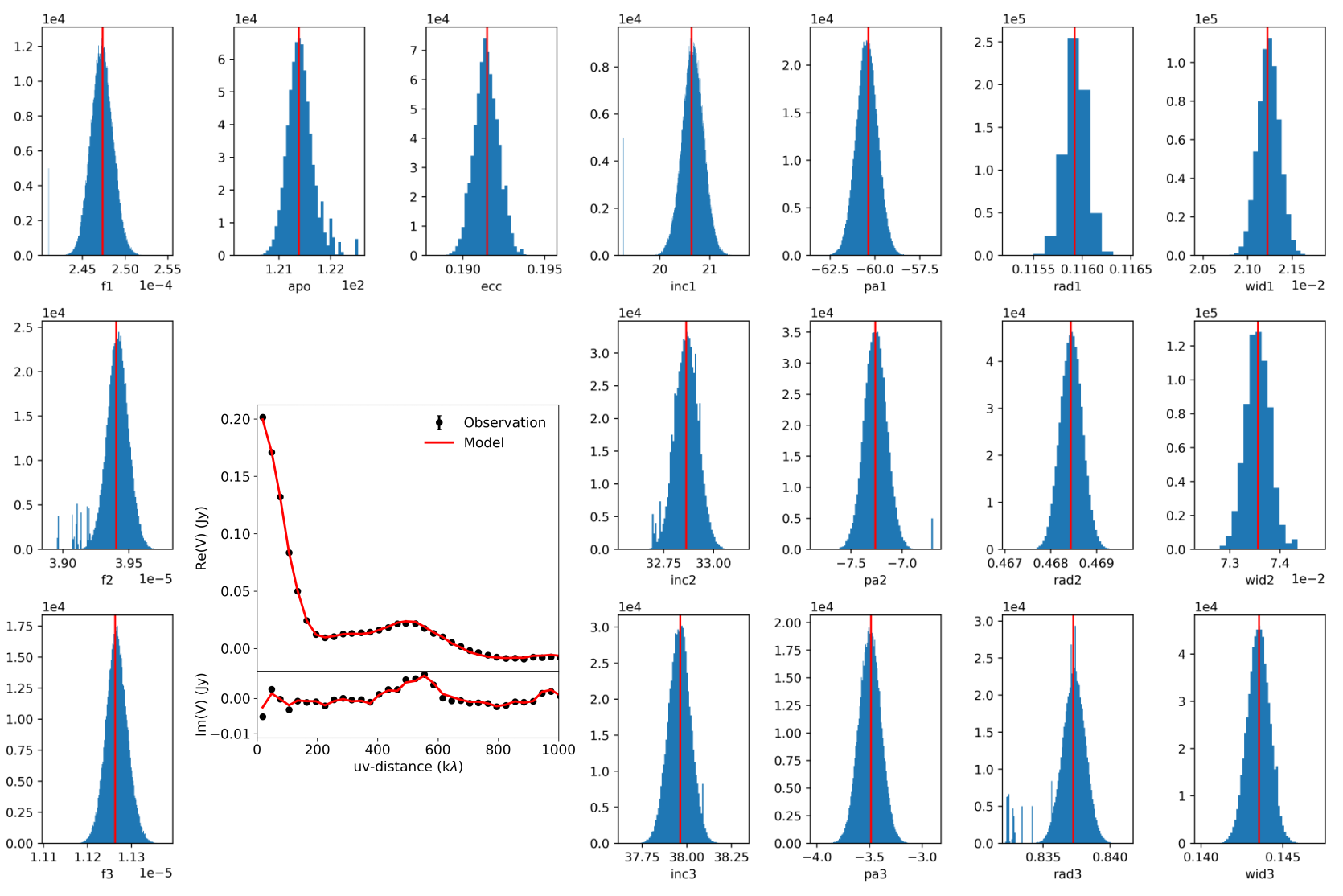}
\figcaption{The quality of MCMC parameter search for Model 3. The \textit{uv}-visibility panel shows the \textit{uv}-plot of both the ALMA observation and its best-fit model, with the upper panel for the real part of visibility and the lower panel for the imaginary part. The surrounding panels show the histograms of chains in the MCMC fitting. The best-fit value is taken from the 50\textsuperscript{th} percentile of the distribution (vertical red line). And the negative and positive uncertainties are taken from 16\textsuperscript{th} and 84\textsuperscript{th} percentile, respectively. The seven columns are for the peak surface brightness, apoapsis angle (inner ring only), eccentricity (inner ring only), inclination, position angle, radius, and width, from left to right. And the three rows (top-down) are for the inner, middle, and outer ring. The units of parameters in the histograms are the same as those in Table \ref{tab:model3}c.
\label{fig:mcmc}} 
\end{sidewaysfigure}

Figure \ref{fig:mcmc} shows the \textit{uv}-plot and posterior distribution of Model 3. 

\clearpage


\section{Equations for the Time-scale Analysis} \label{app:time}

\subsection{Radial Communication Time-scale} \label{app:tc}

For a disk around a binary system of separation $a_b$, its radial communication time-scale $t_c$ can be estimated by \cite{lubow2018} 
\begin{equation}
t_c \approx \frac{8}{5\Omega_b(h/r)_{\text{out}}}\left(\frac{r_{\text{out}}}{a_{\text{b}}}\right)^{3/2}
\end{equation}
where $r_{\text{out}}$ is the outer radius within which the disk is in good radial communication, and $(h/r)_{\text{out}}$ is the aspect ratio at $r_{\text{out}}$, calculated based on the estimated temperature in the disk (Equation \ref{eq:tdust}). The radial communication time-scale of the entire disk (i.e., $r_{\text{out}}$ = 1300 AU) is estimated to be $\sim$ 0.06 Myr.


\subsection{Nodal Precession Time-scale} \label{app:tn}

If there is no radial communication in the disk, each part of the disk shall undergo differential precession with its local precession angular frequency $\omega_{\text{n,local}}$ given by \cite{smallwood2019} 
\begin{equation} 
\omega_{\text{n,local}}=k\left(\frac{a_{\text{b}}}{r}\right)^{7/2}\Omega_{\text{b}},
\end{equation}
where 
\begin{equation}
k=\frac{3}{4}\sqrt{1+3e_{\text{b}}^2-4e_{\text{b}}^4}\,\frac{M_1M_2}{(M_1+M_2)^2}
\end{equation}
is a constant depending on the eccentricity of binary’s orbit $e_{\text{b}}$, and the primary and secondary mass of the binary $M_1$ and $M_2$. However, for a protoplanetary disk in the bending-wave regime, where radial communication is active and prompt, the disk parts at different radii shall undergo \textit{global} precession with the angular frequency $\omega_{\text{n,global}}$ given by \cite{smallwood2019}
\begin{equation} \label{eq:global}
\omega_{\text{n,global}}=k\,\bigg\langle\left(\frac{a_{\text{b}}}{r}\right)^{7/2}\bigg\rangle\,\Omega_{\text{b}},
\end{equation}
and 
\begin{equation}
\bigg\langle\left(\frac{a_{\text{b}}}{r}\right)^{7/2}\bigg\rangle=\frac{\int^{r_{\text{out}}}_{r_{\text{in}}}\Sigma r^3\Omega(a_{\text{b}}/r)^{7/2}\text{d}r}{\int^{r_{\text{out}}}_{r_{\text{in}}}\Sigma r^3\Omega\text{d}r}
\end{equation}
is the angular momentum weighted averaging term, in which $\Omega(r)$ is the angular frequency at a given radius \textit{r}, $\Sigma(r)$ is the disk surface density with a radial dependence of $r^{-3/2}$, and $r_{\text{in}}$ and $r_{\text{out}}$ are inner and outer radius of the disk. Assuming $r_{\text{in}}$ = 32 AU and $r_{\text{out}}$ = 1300 AU, we take $t_{\text{n,global}}$ = 2$\pi$ / $\omega_{\text{n,global}}$ and estimate the global precession time-scale of the entire GW Ori disk to be $\sim$ 0.83 Myr. 


\subsection{Alignment Time-scale} \label{app:ta}

The alignment time-scale $t_{\text{a}}$ is given by \cite{lubow2018} and \cite{bate2000}
\begin{equation}
t_{\text{a}}=\frac{(h/r)^2\,\Omega_{\text{b}}}{\alpha\,\omega_{\text{n}}^2},
\end{equation}
where $\alpha$ = 0.01, \textit{h/r} is defined in Equation \ref{eq:hr}, and the angular frequency of global precession (Equation \ref{eq:global}) is used for $\omega_{\text{n}}$. 


\end{document}